\documentclass{jltp}

\title{Andreev Reflection and Proximity effect}
\author{B. Pannetier and H. Courtois}
\address{C.N.R.S.-Centre de Recherches sur les Tr\`es Basses 
Temp\'eratures,\\ Laboratoire associ\'e \`a l'Universit\'e Joseph Fourier,\\
25 Av. des Martyrs, 38042 Grenoble cedex 9, France}
\runninghead{B. Pannetier and H. Courtois}{Andreev Reflection and Proximity effect}
\input epsf
\epsfverbosetrue

\begin{document}

\maketitle
\begin{abstract}
The Andreev Reflection is the key mechanism for the superconducting
proximity effect. It provides phase 
correlations in a system of non-interacting electrons at 
mesoscopic scales, i.e. over distances much larger than 
the microscopic lengths : Fermi wavelength and elastic electron mean free 
path. This field of research has attracted an increasing interest in 
the recent years in part because of the tremendous development of 
nanofabrication technologies, and also because of the richness of the 
involved quantum effects. In this paper we review some recently achieved 
advances. We also discuss new open questions, in particular 
non-equilibrium effects and proximity effect in systems with ferromagnetic 
elements.

\end{abstract}

\section{Introduction}
The proximity effect is the occurrence of 
superconducting-like properties in non-superconducting materials 
placed in electrical contact with a superconductor (S). It was understood 
in the sixties that superconducting correlation could extend over a 
large length scale in a normal metal (N), even in the absence of attractive 
electron-electron interactions.\cite{Parks,Wolf,Gilabert} 
It is instructive to see that many of the basic phenomena 
in particular regarding tunneling spectroscopy were understood 
although the mesoscopic language was not used. However the 
technology did not allow yet to really manufacture 
complex normal-superconducting circuits in a controllable way 
at this scale. Most experiments were carried out on thin film layers 
and some key experiments were not accessible. 
The two main difficulties - patterning at submicron 
scale and control of S-N interfaces - have been overcome recently 
when state-of-the-art nanofabrication technologies became available in research 
laboratories. In the last ten years a large variety of 
normal-superconductor hybrid structures including phase-sensitive 
devices has been studied, leading to a 
clarification of the basic mechanism as well as a good understanding 
of many non-intuitive observations. This understanding strongly 
benefited from the progresses in mesoscopic physics which called 
the attention to phase-coherent phenomena in disordered 
systems.\cite{Karlsruhe94,Moriond,Comments,Curacao,Superlattice}

It is worth noticing that the important length and energy scales 
in the proximity effect are the same as those governing quantum 
transport or thermodynamic effects in mesoscopic systems. The 
Thouless energy or "correlation energy" $E_{c}$ defined as $\hbar/\tau_{D}$ in 
a disordered metal is of central importance since it characterizes the 
time $\tau_{D}=L^{2}/D$ required by a single electron to feel the sample 
boundary. Here $\hbar$ is the Planck constant, $L$ is the sample length and $D$ is the diffusion 
coefficient. As will be seen, this energy scale plays in some sense 
the role of the energy gap in a non-interacting metal in proximity contact 
with a superconductor. The fact that the actual energy gap is determined 
not by a pairing interaction as in a BCS superconductor but by the 
diffusion of single quasiparticles in the normal metal is a simple result of the remote 
interaction felt by the electronic state when entering the device.

The role of the Andreev reflection\cite{Andreev} is central to the proximity effect 
since it provides the elementary mechanism for converting single 
electron states from a normal metal to Cooper pairs in the 
superconducting condensate. As will be illustrated in this review, the 
actual proximity effect is the result of an interplay between Andreev 
reflection at the N-S interface and long-range coherence in the 
normal metal.

This paper is organized as follows. We first recall the main 
properties of Andreev reflection. We then 
illustrate some consequences on the conductance of the N-S 
junction and of the normal metal itself. Both the sub-gap conductance 
of a tunnel junction and the re-entrance effect are now well understood 
at least in the case of non-interacting metals. In the subsequent 
paragraphs we focus on recently addressed situations which are now 
open topics including non-equilibrium phenomena, thermopower, shot 
noise, density of states and  proximity effects in ferromagnetic 
metals.

\section {The Andreev Reflection}

The Andreev Reflection provides a conversion of the dissipative electrical 
current in the normal metal into a dissipationless supercurrent. The 
signatures of the Andreev Reflection are :
\begin{enumerate}
	\item A two-electron process : Because of the existence of an energy 
	gap at the Fermi energy in the density of states of the 
	superconductor, the transfer of single quasiparticle states with an 
	energy $\epsilon$ below the gap $\Delta$ is forbidden. This situation 
	holds when one restricts oneself to the first-order process as it is the 
	case for an opaque tunnel barrier. Indeed the Giaver tunneling 
	experiments demonstrated that the tunnel conductance of a N-I-S 
	junction directly probes the density of states of a 
	superconductor.\cite{Tinkham} 
	However, another type of transfer is possible when second order processes 
	are allowed. An incoming electron can be transferred into the 
	superconductor if a second electron is also transferred through the interface 
	thus forming a Cooper pair into the superconductor. In terms of 
	single excitations, this process is equivalent to the reflection of 
	a hole.
	
	The consequences of Andreev reflection on the 
	current voltage characteristics of a S-N junction were studied in detail 
	in the so-called BTK theory.\cite{BTK} The barrier strength was 
	characterized by a simple parameter $Z$ ranging from $0$ for a 
	perfect metallic contact to $\infty $ for a low transparency 
	tunnel barrier. With this definition, the transparency reads 
	$t=1/(1+Z^{2})$. The Andreev process is significant when the 
	transparency of the barrier is high. For a perfect contact ($Z=0$) 
	the sub-gap conductance was found to be twice the normal state 
	conductance thus demonstrating the double charge transfer.
	
	\item  Retro-reflection : This curious feature was noticed by Andreev in 
	his original paper on thermal properties of the intermediate state 
	of superconductors. It was observed in particular by Benistant et 
	al.\cite{Benistant} in an elegant experiment with a pure silver single 
	crystal showing that all three components of the velocity changed 
	sign upon reflection on a superconducting interface. The Andreev 
	reflection is a perfect retro-reflection only for electrons 
	incident at the Fermi energy.\cite{Blom} When the energy is above 
	the Fermi energy, the incident 
	electron $(E_{F}+\epsilon, k_{F}+\delta k/2)$ and the reflected hole 
	$(E_{F}-\epsilon, -k_{F}+\delta k/2)$ have different wavelengths in 
	the normal metal. The wavevector mismatch is linear in energy : 
	$\delta k=2 \epsilon / \hbar v_{F}$. In a purely balistic system 
	this results into resonance effects.\cite{Rowell,StJames}
	
	\item  Coherence properties :  The most important property for the
	proximity effect is the phase coherence of the process. The reflected 
	hole carries information 
	both on the phase of the electron state and on the macroscopic phase 
	$\Phi$ of the superconductor. Let us assume that the pair potential 
	is fixed and given by $\Delta e^{i\Phi}$. For a state with energy 
	$\epsilon$ above the Fermi energy the phase change can be written as : 
	$\delta \phi= \Phi + \arccos{(\epsilon/\Delta)}$. One can see 
	that the Andreev reflection of a state at the Fermi energy, 
	$\epsilon = 0$, is accompanied by a phase shift of $\pi/2$. The 
	influence of this phase shift on the resistance of a N-S junction 
	and the difference between Andreev reflection on a superconducting 
	interface and optical reflection on a phase conjugated mirror was 
	recently discussed by C. Beenakker.\cite{Beenakker99} The $\pi/2$ phase 
	shift is at the origin of the finite resistance of a diffusive N-S 
	junction at zero temperature.
	
	In a diffusive metal, the phase shift leads to a loss of interferences 
	beyond an energy-dependent coherence length given by $L_{\epsilon}=\sqrt{\hbar 
	D/\epsilon}$. This mesoscopic length characterizes how far the two electrons 
	from a Cooper pair leaking from the superconductor will diffuse in 
	phase in the normal metal. It appears naturally in the propagation 
	equation for the pair amplitude (Usadel equation). The ultimate 
	cut-off\cite{Lindelof} is the single electron phase memory 
	length $L_{\phi}$ as for the weak localization effect. 
	
	\item  Role of impurities : As far as phase-breaking events 
	can be ignored the presence of impurities does not suppress the 
	quantum interference effects. On the contrary, the diffusion on 
	impurities provides a mechanism to re-direct the trajectories to the 
	interface, therefore enhancing the tranfer at the interface. This 
	is the coherent multiple scattering effect whose importance was 
	emphasized by van Wees et al. in Ref. [\cite{van Wees}]. The superposition of
	multiple coherent transfers through the interface in presence of disorder 
	is at the origin of the so-called reflectionless tunneling.\cite{Beenakker} 
	Namely, the superposition of many second order processes add up to give 
	a first order process of much larger amplitude. In other words, the presence 
	of disorder (in practice confinement) in the normal metal 
	results in a strong enhancement of the conductance of the junction.
    This enhancement is suppressed by phase-breaking effects such as 
	inelastic processes or external magnetic field. Because of the very 
	special relationship between trajectories of 
	electrons and holes, it is possible that some of the phase-breaking 
	processes which are operant in weak localization or Aharonov-Bohm 
	effects where trajectories are well separated might have a smaller 
	dephasing effect here.\cite{Imry}
	
	\item Andreev reflection vs Cooper pair transfer : The 
	Andreev reflection of an electron (or a hole) is 
	equivalent to the transfer of single Cooper pairs in (or out) of the 
	superconducting condensate. The proximity effect is due to the 
	presence of Cooper pairs leaking into the normal metal. The way the pair 
	density builds up in the normal metal is strongly influenced by the 
	presence of impurities, tunnel barrier or boundaries.
    In the theory of non-equilibrium superconductivity\cite{LO} 
	the presence of Cooper 
    pairs in the normal metal is described by the anomalous Green 
	function $F^{R}(x,\epsilon ) = -i \sin{\Theta (x,\epsilon)}$, 
	conveniently expressed through a proximity angle $\Theta (x,
	\epsilon)$ which is a complex function of both the position $x$ and the 
	energy $\epsilon$.

	\item  Role of spin : In the classical Andreev reflection between 
	a pure metal and a BCS superconductor the spin degree of freedom can 
	be ignored : The condensate is formed of Cooper pairs with opposite 
	spins and the spin up and spin down bands of electrons in the N-metal 
	are identical. The picture is the following : An incident electron 
	with spin up (down) is tranferred into the superconductor together 
	with a second electron with spin down (up) to form a Cooper 
	pair. The reflected hole has a spin up (down) since it is associated with 
	a missing down (up) electron. As we will see later, the situation is 
	strongly altered in ferromagnets where the energy bands are 
	spin-dependent.	
\end{enumerate}

\section{The subgap conductance of a N-S interface}

The Andreev reflection describes the elementary microscopic processes 
that occur at an ideal N-S boundary. The description of proximity 
effect requires additional ingredients that are not included in 
the simple models of the interface.\cite{BTK} The zero 
bias anomaly observed in a 
semiconductor-superconductor junction by Kastalsky et al.\cite{Kastalsky} 
could only be understood by taking into account the multiple 
coherent scattering\cite{van Wees} on the impurities which brings back 
the electrons and holes on the interface with the result of increasing 
the effective conductance. This non-local coherent 
effect is generic to the proximity effect. The effect of disorder is 
dramatic. Hekking and Nazarov\cite{Hekking94} 
predicted that the tunnel conductance $G_{T}$ should be enhanced 
by a factor proportional to $G_{T}/G_{N}$, $G_{N}$ being the metallic 
conductance of the N metal in the vicinity of the junction. 
The scattering matrix theory\cite{van Wees,Beenakker,Lesowik,Lambert} 
which extends the Landauer-Buttiker approach of mesoscopic transport 
to include the Andreev reflection at a superconducting interface 
provides a good description of the effect of multiple scattering.
This effect was implicitly taken into account in the 
older quasiclassical theory based upon the Larkin-Ovchinnikov theory 
of non-equilibrium superconductivity.\cite{LO,Volkov,Spivak,Nazarov} 
In many practical situations (disordered metals) the latter approach leads 
to a simple circuit equation for the energy-dependent proximity angle 
$\Theta(\epsilon,x)$.\cite{Nazarov94,Esteve}

The phase-sensitivity was demonstrated by Pothier et al.\cite{Pothier} 
in a loop shaped N-S 
metallic circuit having two N-S junctions in parallel. As in 
Ref. [\cite{Kastalsky}], the subgap conductance showed a strong 
enhancement of the Andreev conductance near zero bias. 
Interestingly, the conductance was periodic with respect to 
the magnetic flux in the superconducting loop. This experiment clearly 
illustrates the existence of an interference effect between the two N-S 
junctions and therefore demonstrates the sensitivity of the Andreev current 
to the superconducting phase. Further recent theoretical works show 
that this phase sensitivity can provide useful informations on 
the quantum fluctuation of the phase of the superconducting island 
in a superconducting transistor.\cite{Hekking95,Feigelman}

\section{The spectral conductance and the re-entrance effect}

As discussed above, the conductance of a N-S junction is enhanced by 
the coherent multiple scattering due to the disorder in the metal. 
The proximity effect can also be directly observed on the 
metallic conductance itself if the conductance of the 
metal is much smaller than the barrier conductance or if the 
design is such that the N-S interface is not in the measured 
circuit. A serie of experiments on Aharonov-Bohm loop 
circuits\cite{Courtois96,Petrashov93} and Andreev 
interferometers\cite{Petrashov93,Dimoulas} have 
elucidated the length and temperature dependences of the phase 
sensitive contribution to the conductance.

It is well established now that the metallic conductance of a normal 
metal N in proximity with a superconductor S exhibits a non-monotonic 
temperature and voltage dependence with a maximum at $kT$ or $eV \approx E_{c}$
(or equivalently $\sqrt{\hbar D/k_{B}T}$ or $\sqrt{\hbar D/eV} \approx L)$. Here $L$ is the length of the 
normal metal and $E_{c}=\hbar D/L^{2}$ is the Thouless energy. This 
so-called re-entrance effect first recognized in Ref. [\cite{Charlat}] 
was also observed in a variety of N-S systems including doped 
semiconductors\cite{Sanquer} and two dimensional 
electron gases connected to a superconductor.\cite{ReentVWees} In the latter  case, the 
position of the maximum could be controlled in-situ by changing the 
diffusion coefficient by an external gate voltage.\cite{Toyoda}

The physics of the re-entrance effect now is well 
understood\cite{Nazarov,Artemenko,Volkov-Lambert,WilhNS} in the case of non 
interacting electrons. The point is that one can define a 
energy-dependent spectral conductance\cite{Charlat} for the electron transport 
through the structure. The measured conductance is the convolution of 
this spectral conductance with the electron energy distribution function.
Interestingly, in both N-S and S-N-S devices the 
temperature $kT=E_{c}$ is a crossover temperature below which the 
conductance diverges in the S-N-S junction (Josephson short circuit) 
or returns to the normal state resistance in the N-S junction 
(re-entrance effect).

At temperatures $kT > E_{c}$ larger than the Thouless energy, the 
magnetoconductance oscillations amplitude decays slowly with temperature with a $1/T$ 
power law.\cite{Courtois96} This behaviour is in clear contrast with 
the exponential decay of the Josephson current over $L_{T}$ in a 
S-N-S junction at high temperature. The physical meaning of this 
long-range effect is that even at relatively high temperatures, electrons 
at the Fermi level form pairs which remain coherent over the whole sample 
length. The relative weight of this population is about 
$E_{c}/k_{B}T$, this factor gives the appropriate order of magnitude 
for the long-range correction to the metallic conductance.\cite{HC_Sup}

\section{The Thermopower}

The first recognized consequence of Andreev reflection is the absence 
of energy transfer through the interface. Actually the initial 
motivation of the original Andreev work\cite{Andreev} was aimed at a 
quantitative understanding of thermal conductivity experiments in a 
type I superconductor that is made of alternate layers of normal and 
superconducting domains. However, to-date, most experiments on mesoscopic N-S 
devices have focused on electrical transport. Only recently the attention 
was called to other properties such as thermopower and thermal 
conductivity.\cite{Claughton}

In a homogeneous ordinary metal the thermopower 
is very small because the electrical conductivity is energy independent. 
This is not true for Kondo alloys where the magnetic scattering time 
is strongly dependent on the electron energy. Thus, alloys such as 
AuFe provide useful materials for low temperature thermocouple sensors. 
In a N-S device it is now understood that the spectral conductance is 
strongly peaked at a characteristic energy given by the Thouless energy. 
This is the origin of the  re-entrance effect discussed above.
Accordingly the thermopower is expected to be larger than in a plain 
normal metal.

Recently J. Eom et al.\cite{Eom} succeeded in the measurements of the 
thermopower of mesoscopic Andreev interferometers. The experiment 
consists in imposing a thermal gradient in a Au wire in contact with a 
micron size loop shaped superconducting circuit. Various sample 
designs were realized. The thermopower was obtained from the voltage 
drop between the reservoirs. It oscillates as a function of the 
magnetic field with a fundamental period corresponding to one flux 
quantum in the superconducting loop. This experiment demonstrates the 
phase-dependence of the thermopower in a mesoscopic Andreev 
interferometer.

\section{Noise experiments}

Shot noise is a consequence of current fluctuations in electrical transport 
due to the discrete nature of the carriers. It has been recognized to 
yield information on the statistics of the charge carrier and on the
correlation of their transmission. This makes noise 
experiments an unique tool in the field 
of mesoscopic quantum transport.\cite{Martin} The shot noise in N-S and S-N-S 
junctions has been investigated recently. Using a calibrated dc-SQUID 
set-up Jehl et al.\cite{Jehl} have observed the doubling of current 
shot noise in a low impedance NbAl structure, in agreement with 
the theoretical predictions of de Jong and Beenakker.\cite{JongNoise} 
This enhancement originates from the fact that the current is carried 
in the superconductor by Cooper pairs in units of 2e.

In the case of a S-N-S junction biased above the critical current, 
multiple Andreev reflections (MAR) take place as recognized by the 
subharmonic gap structures in the I-V characteristic. According to 
this mechanism, an electron is Andreev-reflected at one side of the 
junction as a hole which, after diffusion to the other interface 
is Andreev-reflected as an electron and then continue the cycle. 
At each cycle a Cooper pair is transferred and an energy $eV$ is 
gained. The process ends when the quasiparticle reaches the gap 
energy and therefore escape into the superconductor. A possible charge 
transfer mechanism at low voltage ($eV\ll2\Delta$) is the coherent transfer 
of multiple charge quanta $q*=2\Delta/V$. In such a case a strongly 
enhanced shot noise is expected at low voltage.\cite{Averin} Serious 
indications of this enhanced shot noise mechanism have been obtained 
recently by Dieleman et al.\cite{Dieleman} and Hoss et al..\cite{Hoss}
More experiments are needed to check the nature of the effect.

\section{The density of states}

The density of states is a meaningful quantity which has been first 
investigated in the early days of proximity
superconductivity.\cite{Adkins} The samples were made of a tunnel 
barrier in contact of the N-side of a N-S bilayer 
with a highly transparent interface. The recent development of new 
fabrication techniques enabled a new kind of experiment where the 
density of states can be probed at distinct locations.\cite{Gueron} 
The density of states shows a depression at the 
Fermi energy with a characteristic energy of order of 
the Thouless energy. The experiments have been successfully compared to 
calculations from the Usadel equations\cite{Belzig} provided a 
noticeably large spin-flip rate is included. When the normal metal is 
a "closed system" disconnected from any electron reservoirs a true energy 
gap is expected provided that the system is disordered or 
chaotic.\cite{Fraham96} For example in a S-N-S junction where N is a 
small island, a mini-gap is expected with a strong dependence on 
the phase difference between the two superconducting 
electrodes.\cite{WilhNS,Zhou98} In the ballistic case, special trajectories lead 
to low energy sub-gap states which may depend on the 
dimensionality\cite{StJames} or on the shape of the N-metal.\cite{Fraham96}
New techniques such as local spectroscopy with a low-temperature 
Scanning Tunneling Microscope are welcome for getting more insight 
into the spatial and phase dependence of the density of states in N-S 
structures.

\section{The Josephson effect in SNS structures}

Let us consider an ideal S-N-S structure which consists of a normal metal 
N in-between two superconductors S1 and S2 with respective phases $+\Phi/2$ 
and $-\Phi/2$. Because of the confinement induced by the superconducting gap, 
the electron energy levels in the normal metal consist of phase 
dependent Andreev 
bound-states which, as predicted long ago by Kulik,\cite{Kulik} can 
carry a finite supercurrent. The total supercurrent is given by a 
summation over contribution of the current carrying states which all 
depend upon the phase difference $\Phi$ between the two 
superconductors. For a system in thermal equilibrium, the occupation 
probabilities of each state is given by the Fermi-Dirac distribution 
function. When the phase difference is zero, for each bound 
state there is another degenerate bound state 
(time-reversed state) that carries the same current 
in the opposite direction and therefore the total current is zero.

When the phase difference in a S-N-S junction is non zero, the energies 
of these states are 
different giving rise to a finite supercurrent at low temperature. At 
high temperature, since the thermal population of states with opposite 
current is almost the same the supercurrent is suppressed. The 
suppression occurs when $k_{B}T$ is of order of the spacing between 
states with opposite currents. This picture remains valid in
diffusive systems. In this case this spacing is of order of the Thouless 
energy. Measurements of the Josephson critical current in S-N-S long 
junctions\cite{Courtois95} revealed that the characteristic energy 
$e I_{c}/G_{N}$ at low temperature is of the order of the Thouless energy 
of the sample.\cite{Wilhelm95} This shows again that the Thouless energy is 
the relevant energy scale in a diffusive metal in proximity with a 
superconductor. As the sample $L$ is varied, there is a crossover 
between the superconducting gap for short junctions ($L < \xi_{s}$) 
and the Thouless energy for long junctions ($L > \xi_{s}$).\cite{Dubos2}

\section{Effect of non-equilibrium electron distributions}

According to the above description of the Josephson effect, one sees that the suppression of 
the critical current at high temperature results from a compensation 
of spectral currents with large amplitudes and alternating signs. 
This exact compensation only occurs for a true thermal equilibrium 
distribution function. A small deviation from thermal equilibrium 
of the electron distribution in the normal metal should 
result in a large enhancement of the critical current.

A non-equilibrium distribution can be induced into the metal by injection 
of hot electrons in a four terminal device.\cite{Morpugo98} This 
steady state method has been used recently to demonstrate the operation 
of a mesoscopic S-N-S transistor\cite{WilhelmSNS} where the supercurrent 
of the S-N-S junction could be commanded by a control current. These 
ideas have been implemented experimentally to perform energy 
spectroscopy of the supercurrent in a S-N-S Josephson 
junction.\cite{MorpugoGoteborg}  At the 
lowest temperature and for an optimized design of the control reservoirs 
the sign of the supercurrent could even be reversed, resulting in  a 
$\pi$-junction.\cite{Baselmans} 
An enhancement of the supercurrent under non-equilibrium injection 
was demonstrated recently in a three terminal planar device made of 
aluminum on GaAs. This non-equilibrium induced supercurrent was found 
to exceed the equilibrium supercurrent at high temperature.\cite{Kutchinsky}

A non-equilibium distribution can also be obtained dynamically. Let us 
consider for simplicity a voltage-biased S-N-S junction. According to the 
Josephson equation  the phase difference $\Phi=2eVt/\hbar$ 
becomes time-dependent, leading to a fast oscillation of the Andreev 
levels. Several theoretical predictions have been made recently for 
this regime. F. Zhou et al.\cite{Spivak-NEQ} have calculated the 
conductance enhancement due to the dynamical state in the case of a 
long ($L>L_{T}$) junction for which the equilibrium critical current 
is suppressed. Argaman\cite{Argaman} has considered the non-equilibrium 
Josephson effect in the same limit. In a recent experiments, K. W. 
Lehnert et al.\cite{Lehnert} have observed half-integer Shapiro steps 
in a Nb-InAs-Nb junction which persist to high temperatures in the 
absence of a critical current. The observed power law of their 
amplitude, which is 
reminiscent of that of the proximity-induced AB oscillations\cite{Courtois96} 
in a N-S interferometer, was attributed to the combined time-dependent 
contributions of both the distribution function and the spectral 
current. Current-voltage measurements in a mesoscopic diffusive 
metallic S-N-S junction\cite{Dubos} also revealed strong deviations from the 
resistively-shunted junction model which are due to this 
non-equilibrium dynamical effect.

\section{The Meissner effect}

Like a conventional superconductor, a proximity superconductor expulses 
the magnetic flux : it is the Meissner effect.\cite{Parks} A. Mota et al. 
measured a variety of coaxial structures made of a superconducting 
core (Nb) surrounded by a normal metal envelope (Cu, Ag, Au, 
\ldots).\cite{Mota} These quasi-millimetric samples are in the mesoscopic regime 
because they are very pure, so that the coherence length are of the 
order of the sample size. The price to pay is the use of ultra low 
temperatures (about $100 \mu K$) for reaching the interesting regime $L 
\approx L_{T}$.

At intermediate temperatures, the observed diamagnetic behaviour is 
well described by the quasiclassical theory.\cite{Meiss} A puzzling paramagnetic reentrance 
effect\cite{Mota} has been demonstrated to appear at very low 
temperature when the coherence lengthes are of the order of the 
diameter of the structure. These results motivated a number of 
theoretical studies. C. Bruder and Y. Imry propose an interpretation 
in terms of a whispering gallery for electrons which will not "see" 
the N-S interface,\cite{Bruder} whereas Fauch\`ere et al. propose the 
existence of a repulsive interaction in the normal metal 
reponsible for paramagnetic instability at the N-S 
interface.\cite{NSinstabilite} This topic is still an open question.

\section{Ferromagnetic metal-Superconductor systems}

As opposed to noble normal metals, itinerant ferromagnetic metals 
represent an example of a system with strong electron-electron 
interaction leading to an order state of the electron spins.
The Andreev reflection picture is strongly modified because the incoming electron 
and the Andreev reflected hole occupy opposite spin bands.\cite{Demler} An 
immediate consequence is the total suppression of Andreev reflection in a 
fully spin-polarized metal.\cite{JongFS} This effect has been used 
recently to measure the degree of spin polarization (P) of various 
ferromagnetic metals using direct conductance measurements 
through a superconducting point contact.\cite{Soulen} The spin 
polarization is obtained directly from the differential conductance 
using an adaptation of the BTK theory to include spin polarization. 
With increasing spin polarization the subgap conductance drops 
from about twice the normal state conductance for non-polarized metals to  
a small value in the highly polarized metals. Materials with 
spin polarization up to $90\%$ could be characterized this 
way. Similar observations were made using nanofabricated contacts 
between a superconductor and ferromagnetic 
nanoparticles.\cite{Upadhyay} This technique gives an alternative 
method with higher energy resolution than photo-emission and less
material constraints as compared to single particle tunneling 
experiments. Besides this non-perturbative effect, the injection of 
a spin polarized current\cite{Vasko} was shown to suppress of pairing 
in a superconductor.

The possible existence of long-range proximity effects in 
a ferromagnet has been addressed in several puzzling experiments. 
According to the naive expectation, the strong Fermi wavevector mismatch 
between spin up and spin down energy bands should lead to a very 
short effective coherence length inside the ferromagnet. Using the 
simplified view of a Stoner model with an exchange energy $U$ much 
larger than the superconducting gap energy, the coherence length is 
of order of $L_{U}=\sqrt{\hbar D/2\pi U}$ in a 
single domain ferromagnet with an electron diffusion 
coefficient $D$. For transistion metal such as cobalt, this length is 
of order of a few nanometers. This short range effect has been 
investigated intensively in ferromagnetic-superconductor 
multilayers.\cite{Bulaevski,Radovic,Aarts} However, several recent 
experiments\cite{Petrashovferro94,Giordano96,Giroud98} have shown 
strong resistance drops in mesoscopic ferromagnetic structures 
that cannot be understood by the present models. A 
re-entrance effect similar to that observed in non-interacting metals 
(see above) was even observed in Ref. [\cite{Giroud98}]. A characteristic 
length scale of 200 nm was inferred which is compatible with the 
absence of Aharonov-Bohm oscillations mentioned in this work. 

Interesting theoretical predictions have been reported recently. 
Leadbeater et al.\cite{Leadbeater} have investigated the subgap 
conductance in ferromagnetic mesoscopic structures in the case where 
the exchange energy is of order of the superconducting energy gap. 
Fazio et al.\cite{Fazio} have computed the local density of 
states on both sides of the interface. Zhou et al.\cite{Zhou} 
predicted long-range proximity as due to the penetration of 
the triplet part of the superconducting condensate wave function in 
the ferromagnetic metal. This is made possible in the mesoscopic 
regime in presence of 
spin-orbit interaction in the superconductor. Several authors have recently 
stressed an important consequence of the spin polarization which 
actually leads to an opposite effect.\cite{Falko,Jedema} Since the 
total spin of a Cooper pair in the superconductor is zero, there can 
be no spin current across the interface. Because of the spin polarization 
of the current in the ferromagnet there must be a spin accumulation 
near the interface and therefore an enhanced interface resistance.

Obviously these systems call for further investigation in order to 
understand the origin of the unexpected observations. An extensive 
characterization of these hybrid structures at the mesoscopic scale is 
now highly desirable : magnetic domain structure, density of states in F 
and S, presence of Josephson current in SFS structures, etc.. Also the 
determination of the significant characteristic lengths ( phase memory 
length, spin memory length, spin orbit diffusion length, ..) needs 
substantial experimental efforts.

\section{Summary and perspectives}

The recent years have shown the subtle role of the Andreev 
reflection in the phenomena arising in structures made of 
a normal metal in contact with a superconductor. It has been 
understood that the Thouless energy is the characteristic energy and 
that the induced correlations in the normal metal decay only as $1/T$. 
Many aspects including the case of the ferromagnetic metals and the 
non-equilibrium effects remain unsettled.

Andreev reflection is also a instrument for studying new physical 
effects like in atomic contacts where it has been possible to measure 
the transmission of individual channels.\cite{Goffman} Also at the atomic 
scale, the superconducting proximity effect recently observed in single-walled 
nanotubes\cite{Kasumov} raises interesting questions such as the role 
of electron-electron interactions in one-dimensional systems. In the future, the 
consequences of Andreev reflection will likely play a significant role 
in new areas of mesoscopic physics such as the physics of Charge Density Wave (CDW) 
systems. These systems are known to develop a macroscopic phase coherent 
condensed state. Recent experiments on patterned niobium based\cite{Latishev}
CDW crystals or blue bronze thin films device\cite{VDZ} have shown 
novel mesoscopic phenomena.

\section{Acknowledgements}
We benefited of collaborations with P. Charlat, P. Dubos, F. Hekking, 
M. Giroud, K. Hasselbach, Ph. Gandit, D. Mailly and N. Moussy. 
Valuable discussions with P. Monceau, B. Spivak and 
through the TMR network n$^0$ FMRX-CT97-0143 "Dynamics of superconducting 
circuits" are gratefully acknowledged.


\begin{thebibliography}{9}
%Introduction
\bibitem{Parks}G. Deutscher and P. G. de Gennes, in {\it 
Superconductivity}, Vol. 2, R. D. Parks, ed. Marcel Dekker, New York (1969), p. 1005
\bibitem{Wolf}E. L. Wolf, {\it Principle of Electron Tunneling Spectroscopy}, 
Oxford University Press, (1985)
\bibitem{Gilabert}A. Gilabert, Annales de Physique {\bf 2}, 203 (1977)
\bibitem{Karlsruhe94}Proceedings of the NATO ARL on 
{\it Mesoscopic Superconductivity}, ed. F. W. J. Hekking, G. Sch\" {o}n 
and D. V. Averin, Physica B {\bf 203} 3/4 (1994)
\bibitem{Moriond} {\it Coulomb and Interference effects in Small Electronic 
Structures}, ed. C. Glattli, M. Sanquer, J. Tr\^an Thanh V\^an, 
Series Moriond Condensed Matter Physics, Editions Fronti\`eres, 
(1994) ; {\it Correlated Fermions and Transport in Mesoscopic Systems}, ed. Th. 
Martin, G. Montambaux, J. Tr\^an Thanh V\^an, 
Series Moriond Condensed Matter Physics, Editions Fronti\`eres, 
(1996) ; {\it Quantum Physics at Mesoscopic Scales}, ed. C. Glattli, M. Sanquer, 
J. Tr\^an Thanh V\^an, Series Moriond Condensed Matter Physics, 
Editions Fronti\`eres, (1999)
\bibitem{Comments} D. C. Ralph and V. Ambegaokar, Comments Cond. 
Mat. {\bf 18}, 249 (1998)
\bibitem{Curacao} {\it Mesoscopic Electron Transport}, ed. L. L. Sohn, 
L. Kouwenhoven and G. Sch\"{o}n, NATO ASI Series E {\bf 345}, Kluwer Academic 
Publishers (1997)
\bibitem{Superlattice} Special Issue of Superlattices and 
Microstructures on {\it Mesoscopic Superconductivity}, Vol. {\bf 25} 5/6, ed. 
P. F. Bagwell (1999)
\bibitem{Andreev}A. F. Andreev, Sov. Phys. JETP {\bf 19}, 1228 (1964)

%Andreev Reflection
\bibitem{Tinkham}M. Tinkham, {\it Introduction to Superconductivity}, 
Ed. McGraw Hill, 2nd Ed. (1996)
\bibitem{BTK}G. E. Blonder, M. Tinkham and T. M. Klapwijk, Phys. Rev. 
B {\bf 25}, 4515 (1982) ; A. L. Shelankov, JETP Lett. {\bf 32}, 111 (1982) 
\bibitem{Benistant}P. A. M. Benistant, H. van Kempen and P. Wyder, Phys. 
Rev. Lett. {\bf 51}, 817 (1983)
\bibitem{Blom} H. A. Blom, A. Kadigrobov, A. M. Zagoskin, R. I. 
Shekhter, and M. Jonson, Phys. Rev. B {\bf 57}, 9995 (1998).
\bibitem{Rowell}J. M. Rowell and W. L. Mc Millan, Phys. Rev. 
Lett. {\bf 16}, 453 (1966)
\bibitem{StJames}P. G. de Gennes and D. Saint-James, Phys. Letters 
{\bf4}, 151 (1963)
\bibitem{Beenakker99}C. W. J. Beenakker, cond-mat/9909293 ; J. C. J. 
Paasschens, M. J. M. de Jong, P. W. Brouwer and C. W. J. Beenakker, Phys. 
Rev. A {\bf 56}, 4216 (1997)
\bibitem{Lindelof}J. Kutchinsky, R. Taboryski, T. Clausen, C. B. S\o
rensen, A. Kristensen, P. E. Lindelof, J. Bindeslev Hansen,
C. Shelde Jacobsen and J. L. Skov, Phys. Rev. Lett. {\bf 78}, 931 (1997)
\bibitem{van Wees} B. J. van Wees, B. J. P. de Vries, P. Magn\'{e}e and 
T. M. Klapwijk, Phys. Rev. Lett. {\bf 69}, 510 (1992) 
\bibitem{Beenakker} C. W. J. Beenakker, Review of Modern Phys. {\bf 
69/3}, 731 (1997)
\bibitem{Imry}Y. Imry and A. Stern, Semicond. Sci. Technol. {\bf 9}, 
1879 (1994)
\bibitem{LO}A. I. Larkin and Y. V. Ovchinnikov, Sov. Phys. JETP 
{\bf41}, 960 (1975)

% The subgap conductance of a N-S interface
\bibitem{Kastalsky}A. Kastalsky, A. W. Kleinsasser, L. H. Greene, R. 
Bhat, F. P. Milliken, and J. P. Harbison, Phys. Rev. Lett. {\bf 67}, 3026
(1991)
\bibitem{Hekking94} F. W. J. Hekking and Y. Nazarov, Phys. Rev.
Lett. {\bf 71}, 1625 (1993) ; Phys. Rev. B {\bf 49}, 6847 (1994)
\bibitem{Lesowik}G. B. Lesovik, A. L. Fauch\`{e}re and G. Blatter, Phys. 
Rev. B {\bf 55}, 3146 (1997)
\bibitem{Lambert}C. J. Lambert, J. Phys. Condens. Matter {\bf 3}, 
6579 (1991) ; C. J. Lambert and R. Raimondi, J. Phys. Condens. 
Matter {\bf10}, 901 (1998)
\bibitem{Volkov}A. F. Volkov, A. V. Zaitsev and T. M. Klapwijk, Physica 
C {\bf210}, 21 (1993)
\bibitem{Spivak}F. Zhou, B. Spivak, and A. Zyuzin, Phys. Rev. B 
{\bf 52}, 4467 (1995)
\bibitem{Nazarov}Y. Nazarov and T. H. Stoof, Phys. Rev. Lett.
{\bf 76}, 823 (1996) ; T. H. Stoof and Y. Nazarov, Phys. Rev. B {\bf 53}, 14496 (1996)
\bibitem{Nazarov94}Y. V. Nazarov, Phys. Rev. Lett. {\bf 73}, 1420 
(1994) ; Y. V. Nazarov, in Ref. [\cite{Superlattice}]
\bibitem{Esteve}D. Esteve, H. Pothier, S. Gu\'eron, N. O. Birge, M. 
H. Devoret, in Ref. [\cite{Curacao}]
\bibitem{Pothier}H. Pothier, S. Gu\'{e}ron, D. Est\`{e}ve and M. H.
D\'{e}voret, Phys. Rev. Lett. {\bf 73}, 2488 (1994)
\bibitem{Hekking95}F. W. J. Hekking, L. I. Glazman, and G. Sch\"{o}n, Phys. 
Rev. B {\bf 51}, 15312 (1995) 
\bibitem{Feigelman} M. V. Feigelman, V. B. Geshkenbein, L. B. Ioffe and 
G. Blatter, this volume

%The spectral conductance and the re-entrance effect
\bibitem{Courtois96}H. Courtois, Ph. Gandit, D. Mailly and B.Pannetier, 
Phys. Rev. Lett. {\bf 76}, 130 (1996).
\bibitem{Petrashov93} V. T. Petrashov, V. N. Antonov, P. Delsing and T.
Claeson, Phys. Rev. Lett. {\bf 70}, 347 (1993) ; J.E.T.P. Lett. 
{\bf 60}, 606 (1994) ; Phys. Rev. Lett. {\bf 74}, 5268 (1995) ;  
V. T. Petrashov, R. Sh. Shaikhaidarov, P. Delsing and T. Claeson, JETP 
Lett. {\bf 67}, 513 (1998)
\bibitem{Dimoulas}A. Dimoulas, J. P. Heida, B. J. van Wees, T. M. 
Klapwijk, V. van de Graaf and G. Borghs, Phys. Rev. Lett. {\bf 74}, 602 (1995)
\bibitem{Charlat}P. Charlat, H. Courtois, Ph. Gandit, D. Mailly, A.F. 
Volkov and B. Pannetier, Phys. Rev. Lett. {\bf 77}, 4950 (1996) ; 
H. Courtois, P. Charlat, Ph. Gandit, D. Mailly and B. Pannetier, 
J. Low Temp. Phys. {\bf 116}, 187 (1999) 
\bibitem{Sanquer}W. Poirier, D. Mailly and M. Sanquer, Phys. Rev. 
Lett. {\bf 79}, 2105 (1997)
\bibitem{ReentVWees} S. G. den Hartog, C. M. A. Kapteyn, B. J. van 
Wees, T. M. Klapwijk and G. Borghs, Phys. Rev. Lett. {\bf 77}, 4954 
(1996)
\bibitem{Toyoda}E. Toyoda, H. Takayanagi and H. Nakano, Phys. Rev. B
{\bf 59}, 11653 (1999)
\bibitem{Artemenko}  S.N. Artemenko, A.F. Volkov, A.V. Zaitsev,
Sol. St. Comm. {\bf 30}, 771 (1979)
\bibitem{Volkov-Lambert}A. F. Volkov, N. Allsopp and C. J. Lambert, J. 
Phys. Condens. Matter {\bf 8},  L45 (1996)
\bibitem{WilhNS} A. A. Golubov, F. K. Wilhelm, and A. D. Zaikin, 
Phys. Rev. B {\bf 55}, 1123 (1997)
\bibitem{HC_Sup}H. Courtois, Ph. Gandit, B. Pannetier and D. Mailly 
in Ref. [\cite{Superlattice}]

%Thermopower
\bibitem{Claughton}N. R. Claugton and C. J. Lambert, Phys. Rev. B {\bf 53}, 
6605 (1996)
\bibitem{Eom}J. Eom, C. J. Chien and V. Chandrasekhar, Phys. Rev. 
Lett. {\bf 81}, 437 (1998)

%Noise
\bibitem{Martin}Th. Martin, Phys. Lett. A {\bf 220}, 137 (1996)
\bibitem{Jehl}X. Jehl, P. Payet-Burin, C. Baraduc, R. Calemzuk and M. 
Sanquer, Phys. Rev. Lett. {\bf 83}, 1660 (1999) ; X. Jehl, PhD 
thesis, Grenoble (1999), unpublished.
\bibitem{JongNoise}M. J. M. de Jong and C. W. J. Beenakker, Phys. Rev. 
B {\bf 49}, 16070 (1994)
\bibitem{Averin} Y. Naveh and D. V. Averin, Phys. Rev. Lett. 82, 4090 
(1999)
\bibitem{Dieleman}P. Dieleman, H. G. Bukkems, T. M. Klapwijk, 
M. Schicke and K. H. Gundlach, Phys. Rev. Lett. {\bf 79}, 3486 (1997)
\bibitem{Hoss} T. Hoss, C. Strunk, T. Nussbaumer, R. Huber, U. 
Staufer and C. Schoenenberger, cond-mat/9901129

%density of states et Josephson
\bibitem{Adkins}S. M. Freaks and C. J. Adkins, Phys. Lett. {\bf 7}, 382 
(1969) ; see also Ref. [\cite{Gilabert}]
\bibitem{Gueron} S. Gu\'eron, H. Pothier, N. O. Birge, D. Esteve and M. 
H. Devoret, Phys. Rev. Lett. {\bf 77}, 3025 (1996)
\bibitem{Belzig} W. Belzig, C. Bruder and G. Sch\"{o}n, Phys. Rev. B {\bf 
54}, 9443 (1996)
\bibitem{Fraham96}K. M. Frahm, P. W. Brouwer, J. A. Melsen and C. W. J.
Beenakker, Phys. Rev. Lett. {\bf 76}, 2981 (1996)
\bibitem{Zhou98}F. Zhou, P. Charlat, B. Spivak and B. Pannetier, J. Low 
Temp. Phys. {\bf 110}, 841 (1998)
\bibitem {Kulik}I. O. Kulik, Sov. Phys. JETP {\bf 57}, 1745 (1969), see 
also J. Bardeen and J. L. Johnson, Phys. Rev. B {\bf 5}, 72 (1972)
\bibitem{Courtois95}H. Courtois, Ph. Gandit, and B. Pannetier, Phys. 
Rev. B {\bf 52}, 1162 (1995)
\bibitem{Wilhelm95}F. K. Wilhelm, A. D. Zaikin and G. Sch\"{o}n, J. Low 
Temp. Phys. {\bf 106}, 305 (1997)
\bibitem{Dubos2}P. Dubos, H. Courtois, B. Pannetier, F. K. Wilhelm, A. D. Zaikin 
and G. Sch\"{o}n, unpublished

%Non-equilibrium
\bibitem{Morpugo98}A. F. Morpugo, T. M. Klapwijk and B. van Wees, Appl. 
Phys. Lett. {\bf 72}, 966 (1998)
\bibitem{WilhelmSNS}F. K. Wilhelm, G. Sch\"{o}n and A. D. Zaikin, Phys. 
Rev. Lett. {\bf 81}, 1682 (1998)
\bibitem{MorpugoGoteborg}A. F. Morpugo, this volume
\bibitem{Baselmans}J. J. A Baselmans, A. F. Morpugo, B. J. van Wees and T. M. 
Klapwijk, Nature {\bf 397}, 43 (1999)
\bibitem{Kutchinsky}J. Kutchinsky, R. Taboryski, C. B. S\o rensen, 
J. Bindslev Hansen and P. E. Lindelof, Phys. Rev. Lett. {\bf 83}, 
4856 (1999)
\bibitem{Spivak-NEQ}F. Zhou and B. Spivak, JETP Lett. {\bf 65}, 369 
(1997)
\bibitem{Argaman} N. Argaman, in Ref. [\cite{Superlattice}]
\bibitem{Lehnert}K. W. Lehnert, N. Argaman, H. R. Blank, K. C. Wong, 
S. J. Allen, E. L. Hu, and H. Kroemer,  Phys. Rev. Lett. {\bf 82}, 1265  (1999)
\bibitem{Dubos}F. K. Wilhelm, G. Sch\"{o}n, A. D. Zaikin, A. A. Golubov, 
P. Dubos, H. Courtois, and B. Pannetier, Proceedings of LT22 
(Helsinki, Finland, August 4-11, 1999), to appear in Physica B

% Meissner
\bibitem{Mota}P. Visani, A.C. Mota and A. Pollini, Phys. Rev. Lett. 
{\bf 65}, 1514 (1990) ; F. Bernd M\"{u}ller-Allinger and A. C. Mota, 
cond-mat/9904344
\bibitem{Meiss} F.~B.~M\"uller-Allinger, A.~C. Mota, and W.~Belzig, 
Phys. Rev. B {\bf 59}, 8887 (1999) ; W.~Belzig, C.~Bruder, and A.~L.~Fauch\`ere, 
Phys. Rev. B {\bf 58}, 14531 (1998)
\bibitem{Bruder}C. Bruder and Y. Imry, Phys. Rev. Lett. {\bf 80}, 
5782 (1998) ; Comment : A. Fauch\`{e}re, V. Geshkenbein, and G. Blatter, 
Phys. Rev. Lett. {\bf 82}, 1796 (1999) ; Reply : C. Bruder and Y. Imry, 
Phys. Rev. Lett. {\bf 82}, 1797 (1999)
\bibitem{NSinstabilite} A.~L.~Fauch\`ere, W.~Belzig, and G.~Blatter, 
Phys. Rev. Lett. {\bf 82}, 3336 (1999) 

% ferro-supra
\bibitem{Demler}E. A. Demler, G. B. Arnold and M. R. Beasley, Phys. 
Rev. B {\bf 55}, 15174 (1997)
\bibitem{JongFS}M. J. M. de Jong and C. W. J. Beenakker, Phys. Rev. Lett. 
{\bf 74}, 1657 (1995)
\bibitem{Soulen}R. J. Soulen Jr, J. M. Byers, M. S. Osofsky, B. 
Nadgorny, T. Ambrose, S. F. Cheng, P. R. Broussard, C. T. Tanaka, J. 
Nowak, J. S. Moodera, A. Barry, J. M. D. Coey, Science {\bf 282}, 85 (1998)
\bibitem{Upadhyay} S. Upadhyay, A. Palanisami, R. N. Louie, R. A. Burham, 
Phys. Rev. Lett. {\bf 81}, 3247 (1998)
\bibitem{Vasko}V. A. Vasko, V. A. Larkin, P. A. Kraus, K. R. Nikolaev, 
D. E. Grup, C. A. Nordman and A. M. Goldman, Phys. Rev. Lett. {\bf 78}, 
1134 (1997)
\bibitem{Bulaevski}L. N. Bulaevski, A. I. Buzdin, M. L. Kul\'{i}c and 
S. V. Panyulov, Adv. Phys. {\bf 34}, 175 (1985).
\bibitem{Radovic} C. L. Chien and D. H. Reich, J. of Mag. and Mag. 
Mat. {\bf 200}, 83 (1999)
\bibitem{Aarts}J. Aarts, J. M. E. Geers, E. Brueck, A. A. Golubov and 
R. Coehoorn,  Phys. Rev. B {\bf 56}, 2779 (1997)
\bibitem{Petrashovferro94} V. T. Petrashov, V. N. Antonov, S. V. 
Maksimov and R. Sh. Shaikhaidarov, JETP Lett. {\bf 59}, 
551 (1994) ; V. T. Petrashov, I. A. Sosnin, I. Cox, A. Parsons and C. 
Troadec, Phys. Rev. Lett. {\bf 83}, 3281 (1999)
\bibitem{Giordano96}H. D. Lawrence and N. Giordano, J. Phys. Cond. Mat. 
{\bf 8}, L653 (1996)
\bibitem{Giroud98}M. Giroud, H. Courtois, H. Hasselbach, D. Mailly and B. 
Pannetier, Phys. Rev. B {\bf 58}, 11872 (1998)
\bibitem{Leadbeater}M. Leadbeater, C. J. Lambert, K.E. Nagaev, R. Raimondi 
and A. F. Volkov, Phys. Rev. B {\bf 59}, 12264 (1999)
\bibitem{Fazio}R. Fazio and C. Lucheroni, Europhys. Lett. {\bf 45}, 
707 (1999)
\bibitem{Zhou}F. Zhou and B. Spivak, cond-mat/9906177
\bibitem{Falko} V. I. Fal'ko, C. J. Lambert and A. F. Volkov, JETP Lett. 
{\bf 69}, 532 (1999)
\bibitem{Jedema}F. J. Jedema, B. J. van Wees, B. H. Hoving, A. T. Filip, 
T. M. Klapwijk, cond-mat/9901323

%Summar and perspectives
\bibitem{Goffman}E. Scheer, P. Joyez, D. Est\`eve, C. Urbina and M. H. Devoret, Phys. 
Rev. Lett. {\bf 78}, 3535 (1997) ; M. F. Goffman, R. Cron, A. Levy Yeyati, P. Joyez, 
M. H. Devoret, D. Est\`{e}ve, C. Urbina, this volume
\bibitem{Kasumov}A. Yu. Kasumov, R. Deblock, M. Kociak, B. Reulet, H. 
Bouchiat, I. I. Khodos, Yu. B. Gorbatov, V. T. Volkov, C. Journet, M. 
Burghard, Science {\bf 284}, 1508 (1999)
\bibitem{Latishev}Yu. I. Latishev, O. Laborde, P. Monceau and S. 
Klaum\"{u}nzer, Phys. Rev. Lett. {\bf 78}, 919 (1997) ; Yu. I. Latishev, 
B. Pannetier and P. Monceau, Eur. Phys. J. {\bf 3}, 421 (1998)
\bibitem{VDZ}H. S. J. van der Zant, O. C. Mantel, C. P. Heij and C. 
Dekker, Synthetic Metals {\bf 86}, 1781 (1997) 

\end{thebibliography}
\end{document}